\documentclass[camera ready]{Interspeech}
\usepackage{threeparttable}
\usepackage{dblfloatfix}

\usepackage{float}

\title{Learning to Attend to Depression-Related Patterns: An Adaptive Cross-Modal Gating Network for Depression Detection}

\author[affiliation={1,2,3},  equalcontribution]{Hangbin}{Yu}
\author[affiliation={1,3}, equalcontribution]{Yudong}{Yang}
\author[affiliation={1,2,3}, correspondingauthor]{Rongfeng}{Su}
\author[affiliation={1,2,3}]{Nan}{Yan}
\author[affiliation={1,2,3}, correspondingauthor]{Lan}{Wang}

\address{
    $^1$  Shenzhen Institutes of Advanced Technology, Chinese Academy of Sciences, China \\
    $^2$ University of Chinese Academy of Sciences \\
    $^3$ Key Laboratory of Biomedical Imaging Science and System, Chinese Academy of Sciences, China
}

\email{hb.yu@siat.ac.cn, rf.su@siat.ac.cn, lan.wang@siat.ac.cn}


\usepackage{comment}

\begin{document}

\maketitle

\setlength{\textfloatsep}{6pt} 
\setlength{\dbltextfloatsep}{6pt} 
\setlength{\floatsep}{6pt}       
\setlength{\textfloatsep}{6pt}   
\setlength{\intextsep}{6pt}

\begin{abstract}
Automatic depression detection using speech signals with acoustic and textual modalities is a promising approach for early diagnosis. Depression-related patterns exhibit sparsity in speech: diagnostically relevant features occur in specific segments rather than being uniformly distributed. However, most existing methods treat all frames equally, assuming depression-related information is uniformly distributed and thus overlooking this sparsity. To address this issue, we proposes a depression detection network based on Adaptive Cross-Modal Gating (ACMG) that adaptively reassigns frame-level weights across both modalities, enabling selective attention to depression-related segments. Experimental results show that the depression detection system with ACMG outperforms baselines without it. Visualization analyses further confirm that ACMG automatically attends to clinically meaningful patterns, including low-energy acoustic segments and textual segments containing negative sentiments.
\end{abstract}

\section{Introduction}
Depression has become one of the most pressing global public health concerns. It is a prevalent mental disorder characterized by persistent low mood, psychomotor retardation, and reduced motivation ~\cite{ferrari2013burden}. Conventional diagnostic procedures primarily rely on patients actively seeking clinical consultation, followed by subjective self-reports and clinician assessments ~\cite{insel2010research}. Such approaches often suffer from delayed diagnosis, limited scalability, and difficulties in continuous symptom monitoring. Therefore, the development of fast and accurate automatic depression detection methods is of significant clinical and social importance. Recently, automatic depression detection methods based on speech signals have attracted increasing attention due to their non-invasive nature and high accessibility.

Speech signals encompass acoustic and textual modalities that provide complementary information for depression detection. Existing studies have primarily focused on extracting depression-related embeddings from each modality and fusing them to enhance diagnostic performance. For example,Al Hanai et al.~\cite{al2018Detecting} pioneered the application of LSTM networks to capture long-term temporal dynamics in both acoustic and textual modalities for depression detection. Their findings revealed that multimodal integration substantially surpasses unimodal approaches, highlighting the complementarity of both modalities. Subsequently, researchers have explored various network architectures for extracting deep depression-related embeddings, such as ECAPA-TDNN \cite{wang2022ecapa}, parallel CNN \cite{yin2023depression}, and Speechformer-CTC \cite{wang2024speechformer}. 

With the advancement of large-scale pre-trained models, an increasing number of researchers have adopted pre-trained speech models and language models for extracting acoustic and textual depression-related embeddings, respectively. Pre-trained speech models (Wav2Vec \cite{schneider2019wav2vec}, HuBERT \cite{hsu2021hubert}, and WavLM \cite{chen2022wavlm}) effectively encode prosodic and paralinguistic information,including pitch variation, speaking rate, and pause patterns,that has been shown to correlate with depressive symptoms \cite{wu2023self,dumpala24b_interspeech}. Meanwhile, pre-trained language models (BERT \cite{devlin2019bert}, RoBERTa\cite{liu2019roberta}, and various LLMs\cite{loweimi25_interspeech}) capture semantic, emotional, and psycholinguistic features, such as negative sentiment and self-referential language, which also constitute typical markers of depression. Existing approaches typically proceed by first extracting frame-level acoustic and textual features using pre-trained models, followed by direct feeding into downstream networks. These features are then compressed into utterance-level embeddings via simple aggregation strategies (such as global average pooling) for depression detection. These methods rest on the assumption that all frames within an utterance contribute equally to depression detection modelling.

However, prior studies~\cite{wang2024speechformer, simantiraki2017glottal} have demonstrated the sparsity of depression-related patterns, implying that detection models should selectively attend to depression-related segments rather than processing entire utterances uniformly. For example, depression-related acoustic patterns exhibit localized concentration in specific temporal-spectral regions~\cite{niu2021time} ,such as diminished pitch modulation, lengthened pauses, and flattened intonation ~\cite{mundt2012vocal}, rather than being uniformly distributed across utterances. This inherent sparsity challenges conventional approaches that assume homogeneous depression-related informativeness across all frames, phonemes, or sub-words. A similar phenomenon is observed in the textual modality: individuals with depression tend to use more negative words and self-referential pronouns ~\cite{edwards2017meta, tackman2019depression}, yet these diagnostically informative patterns usually occupy only limited portions of utterances. Therefore, effectively capturing depression-related patterns necessitates mechanisms capable of identifying and focusing on sparse, diagnostically salient segments while filtering out redundant or irrelevant information. 

To address the sparsity of depression-related patterns in speech signals, this paper proposes a depression detection network based on Adaptive  Cross-Modal Gating (ACMG). \begin{figure*}

\centering 
\includegraphics[width=1\linewidth]{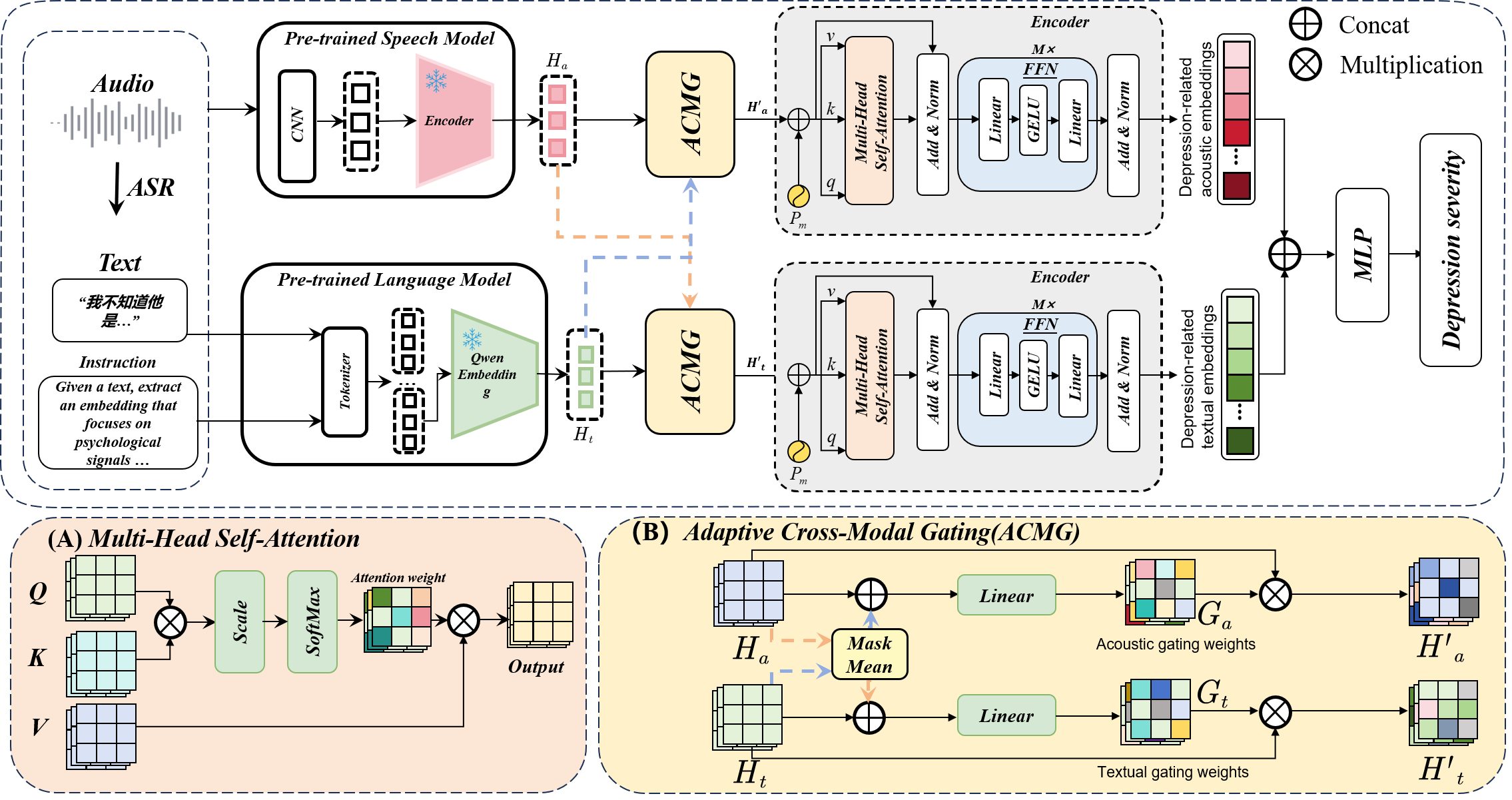} 
\caption{Overall framework of the proposed Adaptive  Cross-Modal Gating (ACMG) network, (A) and (B) illustrate the details of Multi-Head Attention and ACMG modules, respectively.} \label{fig:framework} 

\end{figure*}The ACMG mechanism adaptively reassigns frame-level weights across acoustic and textual modalities, enabling the model to focus on depression-related segments while suppressing noise and irrelevant information. Experimental results demonstrate that the depression detection system with the ACMG mechanism outperforms the baseline system without it. Furthermore, visualization analyses reveal that the introduction of the ACMG mechanism enables the model to automatically attend to depression-related patterns in both modalities, including low-energy segments with prolonged pauses in the acoustic modality, as well as the segments characterized by negative sentiment and first-person pronouns in the textual modality.

\section{Method}

Motivated by the sparse nature of depressive indicators, this paper proposes a depression detection network based on Adaptive  Cross-Modal Gating (ACMG) mechanism. This mechanism selectively enhances clinically significant segments while suppressing neutral or irrelevant content.

\subsection{Overall Framework}

As illustrated in Figure~\ref{fig:framework}, the proposed framework adopts a dual-branch architecture incorporating the ACMG module. The inputs consist of audio data and the corresponding text data, where the latter is obtained through a well-trained ASR model using the WeNet~\footnote{\url{https://github.com/wenet-e2e/wenet}} toolkit. Specifically, the audio is processed by a pre-trained speech model to extract frame-level acoustic features, while the ASR-generated transcript is encoded by an instruction-aware large language model to produce contextualized token representations. Subsequently, the ACMG module performs adaptive cross-modal feature selection by leveraging global context from each modality to refine the other, thereby emphasizing depression-relevant speech frames and textual tokens. The resulting gated representations $\mathbf{H}'_a$ and $\mathbf{H}'_t$ are then modeled by modality-specific Transformer encoders, concatenated, and fed into a MLP classifier to predict the depression severity level.

\subsection{Pre-trained Speech Model}
We employ HuBERT~\cite{hsu2021hubert, Su2026AcousticTextualInconsistency} as the pre-trained speech model,  given its superior performance compared to alternative pre-trained speech models. Furthermore, motivated by findings that intermediate layers of self-supervised speech models encode rich depression-related information~\cite{gat2022speaker,morais2022speech,adoma2020comparative}, we extract representations from the 12th layer of HuBERT as features for depression detection.
The frame-level acoustic features obtained from the audio encoder are denoted as:
\begin{equation}
\mathbf{H}_a = \{ \mathbf{h}_{a}^{1}, \mathbf{h}_{a}^{2}, \dots, \mathbf{h}_{a}^{T_a} \} \in \mathbb{R}^{T_a \times d},
\end{equation}
where $T_a$ and $d$ denote the number of acoustic frames and the hidden dimensionality, respectively.
During training, the parameters of the pre-trained speech model are frozen to preserve the learned acoustic information and prevent overfitting on the limited depression detection data. The resulting frame-level acoustic features are subsequently fed into the ACMG module.

\subsection{Pre-trained Language Model}
We employ Qwen-Embedding-0.6B~\cite{qwen3embedding} as the pre-trained language model to encode transcripts into high-quality semantic representations. Unlike standard pre-trained language models (such as BERT~\cite{devlin2019bert} and RoBERTa~\cite{liu2019roberta}) that generate general-purpose semantic vectors, Qwen-Embedding-0.6B is an instruction-aware large language model specifically designed for controllable representation learning. This capability is particularly crucial for depression detection, where subtle linguistic cues, such as negative affect, self-referential expressions, and cognitive distortions, may be overshadowed by dominant semantic content in generic embeddings.

By incorporating task-oriented instructions during textual feature extraction, the model can condition its representation space on depression-specific objectives, thereby emphasizing psycholinguistic markers relevant to severity assessment. This instruction-guided mechanism aligns the encoding process with the downstream classification task, yielding more discriminative and task-adaptive textual representations. The textual features extracted from the pre-trained language model are denoted as:
\begin{equation}
\mathbf{H}_t = \{ \mathbf{h}_{t}^{1}, \mathbf{h}_{t}^{2}, \dots, \mathbf{h}_{t}^{T_t} \} \in \mathbb{R}^{T_t \times d},
\label{eq:eq1}
\end{equation}
where $T_t$ and $d$ denote the number of textual tokens and the hidden dimensionality, respectively.
\subsection{Adaptive  Cross-Modal Gating (ACMG) Mechanism}
To enable the model to focus on sparse depression-related patterns within an utterance,such as diminished pitch modulation, lengthened pauses, and flattened intonation in speech, alongside negative lexical usage in text,we propose ACMG mechanism.
\subsubsection{Global Context Extraction.}
Given the acoustic representations $\mathbf{H}_a \in \mathbb{R}^{T_a \times d}$ and textual representations $\mathbf{H}_t \in \mathbb{R}^{T_t \times d}$, we first compute global contextual summaries for each modality.
Since audio recordings and transcripts vary in length across samples, zero padding is applied to unify the sequence lengths within a batch for efficient parallel computation. To ensure that padded positions do not contribute to the global representation, we employ masked mean pooling:
\begin{equation}
\setlength{\abovedisplayskip}{4pt} 
\setlength{\belowdisplayskip}{4pt} 
\bar{\mathbf{h}}_a =
\frac{1}{N_a}
\sum_{i=1}^{T_a}
m_a^{(i)} \mathbf{h}_a^{(i)}, 
\quad
\bar{\mathbf{h}}_t =
\frac{1}{N_t}
\sum_{j=1}^{T_t}
m_t^{(j)} \mathbf{h}_t^{(j)},
\label{eq:eq2}
\end{equation}
where $m^{(i)} \in \{0,1\}$ denotes the padding mask, and $N_a$, $N_t$ are the numbers of valid (non-padding) elements. 
As derived from the above equations, since $\mathbf{h}_a^{(i)}$ and $\mathbf{h}_t^{(j)}$ contain local information in acoustic and textual space, respectively, $\bar{\mathbf{h}}_a$ and $\bar{\mathbf{h}}_t$ can represent global depression-related contexts for each modality. These summary vectors are subsequently employed to guide the cross-modal weight reassignment process.
\subsubsection{Adaptive Gating Weights.}
To examine the effect of modality-specific and cross-modal guidance, we introduce two gating weight calculation strategies, namely unimodal and cross-modal gating. 
To compute adaptive gating weights, we first expand the global context vectors to match the temporal dimensions of the features. In unimodal gating, each global vector is replicated along its own modality’s time dimension and concatenated with the corresponding modality’s frames. In cross-modal gating, the textual global context $\bar{\mathbf{h}}_t$ is replicated $T_a$ times to align with acoustic frames, and the acoustic global context $\bar{\mathbf{h}}_a$ is replicated $T_t$ times to align with textual frames. The cross-modal gating weights for both acoustic and textual modalities are  defined as:
\begin{equation}
\setlength{\abovedisplayskip}{4pt} 
\setlength{\belowdisplayskip}{4pt} 
\begin{aligned}
\mathbf{G}_a &= \sigma \Big( W_a [ \mathbf{H}_a \,\Vert\, \tilde{\mathbf{h}}_t ] \Big), &
\mathbf{G}_t &= \sigma \Big( W_t [ \mathbf{H}_t \,\Vert\, \tilde{\mathbf{h}}_a ] \Big)
\end{aligned}
\label{eq:gt_ga}
\end{equation}
where $\tilde{\mathbf{h}}_t \in \mathbb{R}^{T_a \times d}$ and $\tilde{\mathbf{h}}_a \in \mathbb{R}^{T_t \times d}$ denote the temporally expanded textual and acoustic global representations, obtained by replicating $\bar{\mathbf{h}}_t$ and $\bar{\mathbf{h}}_a$ , respectively.
$\parallel$ denotes feature concatenation,
$W_a \in \mathbb{R}^{2d \times 1}$ is a learnable projection matrix, 
and $\sigma(\cdot)$ is the sigmoid activation function. 
For the unimodal gating weights ,they are defined as:
\begin{equation}
\begin{aligned}
\mathbf{G}_a &= \sigma \Big( W_a [ \mathbf{H}_a \,\Vert\, \tilde{\mathbf{h}}_a ] \Big), &
\mathbf{G}_t &= \sigma \Big( W_t [ \mathbf{H}_t \,\Vert\, \tilde{\mathbf{h}}_t ] \Big)
\end{aligned}
\label{eq:gt_gaaa}
\end{equation}
where $\tilde{\mathbf{h}}_t \in \mathbb{R}^{T_t \times d}$ and $\tilde{\mathbf{h}}_a \in \mathbb{R}^{T_a \times d}$  denotes the expanded version of the global context $\bar{\mathbf{h}}_a$ and $\bar{\mathbf{h}}_t$, 
and $W_t \in \mathbb{R}^{2d \times 1}$ is a learnable projection matrix.
\subsubsection{Feature Refinement.}
The computed gating weights are subsequently applied to modulate the contributions of individual frames and tokens, amplifying depression-related pattern while suppressing irrelevant information in both modalities. The new acoustic and textual features are defined as:
\begin{equation}
\mathbf{H}'_a = \mathbf{G}_a \odot \mathbf{H}_a, 
\quad
\mathbf{H}'_t = \mathbf{G}_t \odot \mathbf{H}_t,
\end{equation}
where $\odot$ denotes element-wise multiplication.
This cross-modal gating mechanism facilitates selective enhancement of acoustic frames informed by textual semantics, and reciprocally, highlighting of salient tokens guided by acoustic patterns. By adaptively reweighting features, the model emphasizes diagnostically relevant cues,such as flattened intonation co-occurring with negative self-referential expressions,while suppressing task-irrelevant variations. This mutual refinement process yields more robust representations for depression severity assessment.

\section{Experiment}
\subsection{Datasets}
We conduct experiments on two datasets: the PDCD2025~\cite{Su2026AcousticTextualInconsistency} dataset and the publicly available DAIC-WOZ~\cite{gratch2014distress} benchmark. 

PDCD2025 consists of 22.9 hours of online telephone counseling sessions from 272 participants aged 12–45. The dataset includes healthy controls and clinically diagnosed depressed subjects.
Healthy participants were screened using the Self-Rating Depression Scale (SDS)~\cite{zung1965self} and the Self-Rating Anxiety Scale (SAS)~\cite{zung1971rating} to exclude individuals with subclinical symptoms. Each depressed subject was evaluated using the Hamilton Depression Rating Scale (HAMD) and received a professional diagnosis from certified psychiatrists. Healthy controls had no history of depression or other psychiatric disorders. The dataset comprises three severity labels: Healthy, Mild, and Moderate.

DAIC-WOZ is a widely used benchmark for depression detection, comprising 189 English clinical interviews annotated with PHQ-8 scores. The dataset provides official subject-independent splits for training, development, and testing. Following the standard evaluation protocol, we utilize only participant speech segments, excluding interviewer questions. DAIC-WOZ contains two binary labels: not depressed (PHQ-8 $<$ 10) and depressed (PHQ-8 $\geq$ 10).

\subsection{Experimental Results}

From the results in Table~\ref{tab:HXQ} and Table~\ref{tab:DAICWOZ}, three major trends can be observed.

\textbf{I)} Systems with the ACMG mechanism consistently outperformed their non-ACMG counterparts in subject-level average accuracy. As shown in the last line of Table 1, the optimal configuration employing cross-modal gating weights attains 81.25\% average accuracy, yielding a 1.47\% absolute gain relative to the baseline.
\begin{table}[t]
\centering
\begin{threeparttable}
\setlength{\heavyrulewidth}{1.2pt}
\caption{Performance comparison on PDCD2025 using official 5-fold cross-validation.}
\label{tab:HXQ}
\scriptsize

\begin{tabular}{lcccc}
\toprule
\textbf{Method} & \textbf{Acc.} & \textbf{F1} & \textbf{Prec.} & \textbf{Recall} \\
\midrule
\multicolumn{5}{c}{\textit{Comparison with Existing Methods}} \\
\midrule
Al Hanai et al. (2018)~\cite{al2018Detecting}* 
& 72.43 & 72.46 & 72.25 & 72.67 \\
Gómez-Zaragoza et al. (2025)~\cite{GomezZaragoza2025SpeechTextFM}* 
& 75.37 & 75.55 & 76.03 & 75.07 \\
\midrule
\midrule
\multicolumn{5}{c}{RoBERTa \textit{as Pre-trained Language Model }} \\
\midrule
Transformer(RoBERTa)~\cite{Su2026AcousticTextualInconsistency} 
& 75.74 & 76.17 & 75.64 & 76.70 \\
\quad+ ACMG w Unimodal 
& 77.57 & 77.33 & 77.07 & 77.59 \\
\quad+ ACMG w Cross-modal 
& 78.68 & 78.33 & 78.33 & 78.33 \\
\midrule
\midrule
\multicolumn{5}{c}{Qwen-embedding-0.6B \textit{as Pre-trained Language Model }} \\
\midrule
Transformer(Qwen)\tnote{$\dag$}
& 79.78 & 79.48 & 79.37 & 79.59 \\
\quad+ ACMG w Unimodal
& 80.88 & 80.58 & 80.31 & 80.85 \\
\quad+ ACMG w Cross-modal 
& \textbf{81.25} & \textbf{80.77} & \textbf{80.61} & \textbf{80.93} \\
\bottomrule
\end{tabular}

\begin{tablenotes}
\scriptsize
\item * Results are reimplemented under our experimental settings.
\item $\dag$ The neural network structure follows that of~\cite{Su2026AcousticTextualInconsistency}, with the exception that the pre-trained language model employs Qwen-Embedding-0.6B instead of RoBERTa.
\end{tablenotes}
\end{threeparttable}
\end{table}

\begin{table}[t]
\centering
\begin{threeparttable}
\setlength{\heavyrulewidth}{1.2pt}
\caption{Experimental Results on the DAIC-WOZ Benchmark using Official Development Partition. }
\label{tab:DAICWOZ}
\scriptsize 
\setlength{\tabcolsep}{4pt}
\begin{tabular}{lccccc} 
\toprule
\textbf{Method} & \textbf{Audio} & \textbf{Text} & \textbf{F1} & \textbf{Prec.} & \textbf{Recall}  \\
\midrule
Al Hanai et al. (2018) ~\cite{al2018Detecting} & $\checkmark$ & $\times$ & 50.00 & 71.00 & 38.00 \\
Al Hanai et al. (2018) ~\cite{al2018Detecting} & $\times$ & $\checkmark$ & 59.00 & 71.00 & 50.00 \\
Gómez-Zaragoza et al. (2025) ~\cite{GomezZaragoza2025SpeechTextFM} & $\checkmark$ & $\checkmark$ & 64.30 & - & - \\
Wang et al. (2022) ~\cite{wang2022unsupervised} & $\checkmark$ & $\checkmark$ & 68.34 & - & - \\
\midrule
Transformer(Qwen)  & $\checkmark$ & $\checkmark$ & 64.38 & 64.77 & 64.13 \\
\quad+ ACMG w Cross-modal & $\checkmark$ & $\checkmark$ & \textbf{69.39} & \textbf{71.79} & \textbf{68.48}  \\
\bottomrule
\end{tabular}

\end{threeparttable}
\end{table}

\textbf{II)} In contrast to conventional language models, instruction-guided language model facilitate more effective extraction of depression-related semantic priors, yielding substantial performance gains. For instance, the ``Transformer(Qwen)'' configuration achieves a 4.04\% absolute improvement in average accuracy over the ``Transformer(RoBERTa)'' counterpart on PDCD2025 dataset.

\textbf{III)} The proposed ACMG network exhibits strong generalization capability. Leveraging only official participant speech segments of DAIC-WOZ dataset, our system achieves a strong F1 score of 69.39\%.

Overall, the experimental results confirm that adaptive multimodal gating, cross-modal interaction, and instruction-guided language representations each contribute complementary information, and their integration yields a highly effective system for multimodal depression detection.

\subsection{\textbf{Impact of the Gated Mechanism}}
\subsubsection{Acoustic Gating Analysis} 
To quantitatively evaluate whether the ACMG mechanism facilitates the capture of clinically diagnostic acoustic information, we analyze the correlation between the learned acoustic gating weights and acoustic energy extracted from the eGeMAPS feature set. Specifically, the energy sequence is temporally aligned with the gating outputs, and the Pearson correlation coefficient is subsequently computed to quantify their linear relationship.

The results show a consistent negative correlation across depression severity levels: $-0.263$ (Normal), $-0.342$ (Mild), and $-0.374$ (Moderate), with an overall correlation of $-0.329$. This indicates that the ACMG mechanism assigns higher weights to low-energy speech segments. Furthermore, the increasing magnitude of negative correlation from Normal to Moderate suggests that the gating mechanism progressively strengthens its focus on depression-related acoustic patterns as symptom severity intensifies.

\begin{figure}[t]
    \centering
    \includegraphics[width=\linewidth]{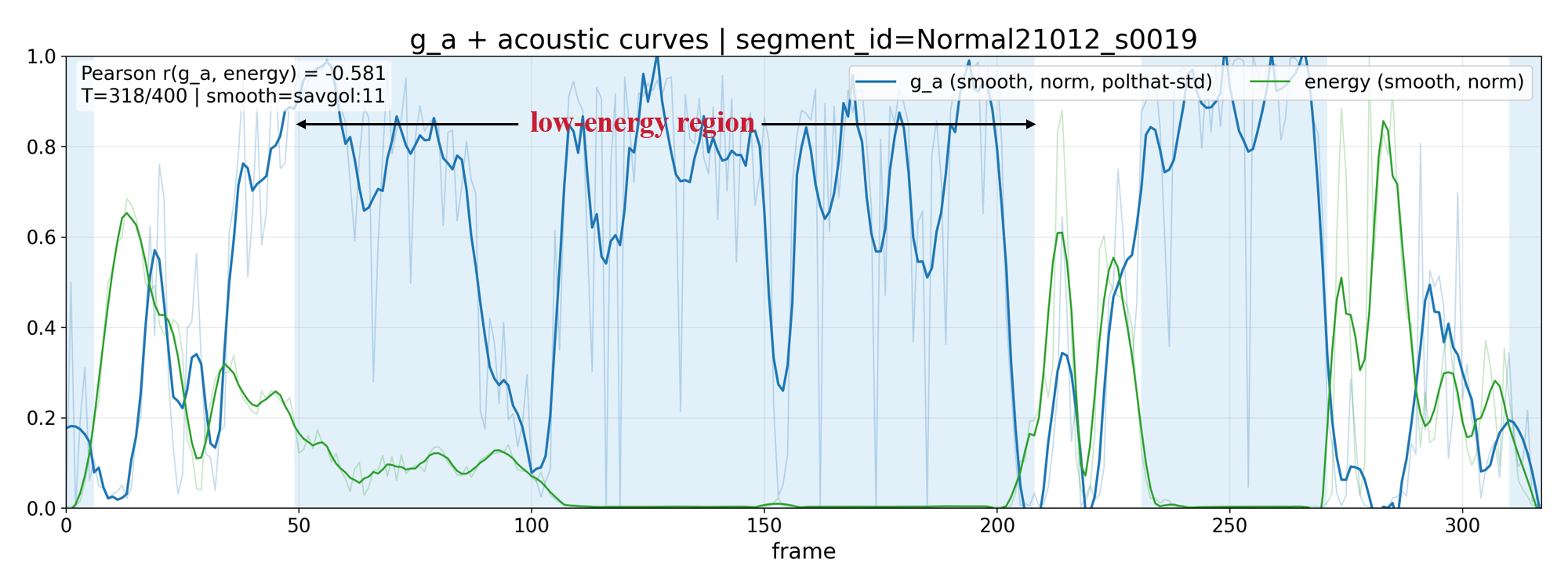}
    \vspace{-4mm}

    \includegraphics[width=0.95\linewidth,height=0.8\linewidth,keepaspectratio]{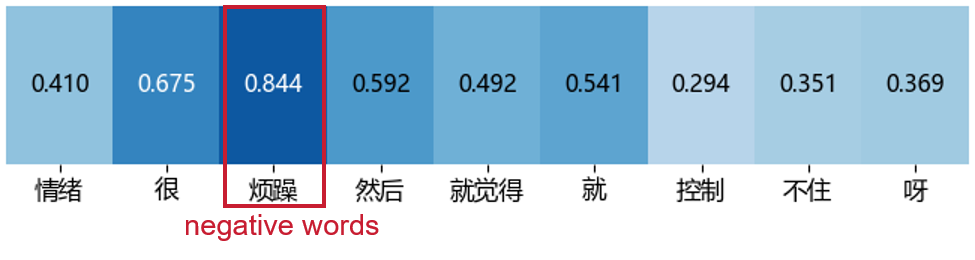}
    \vspace{-2mm}

    \includegraphics[width=\linewidth]{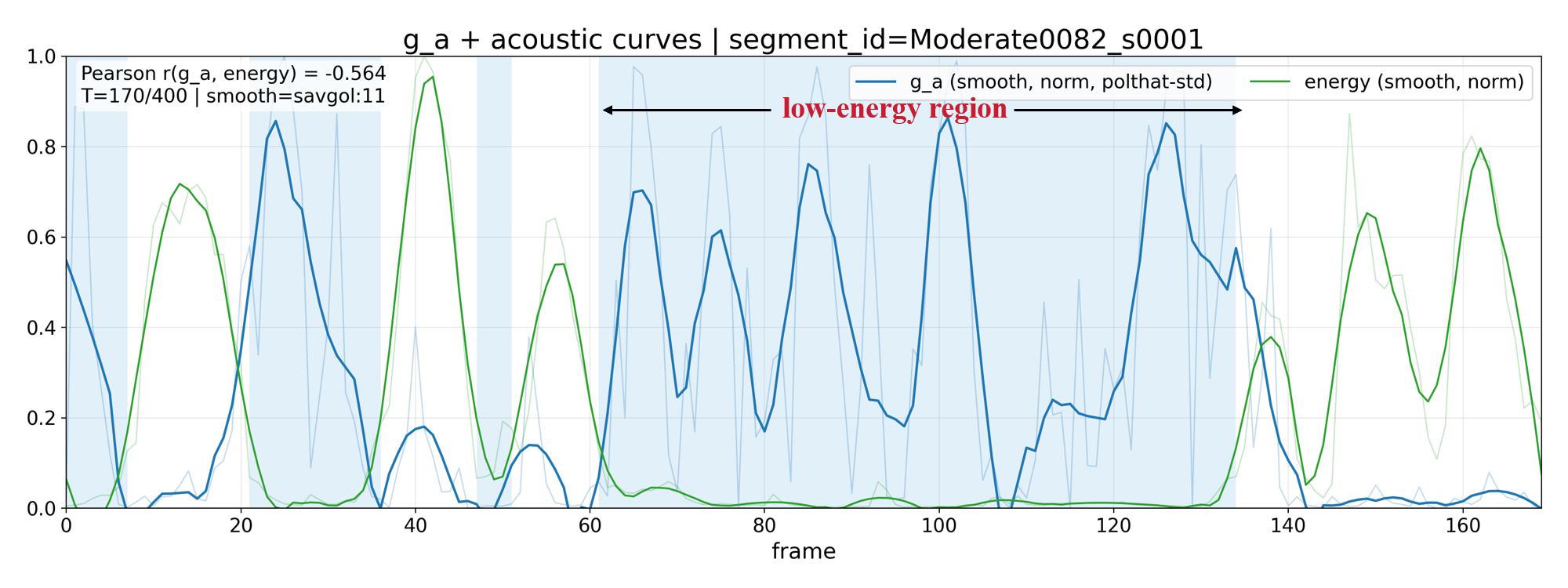}
    \vspace{-4mm}

    \includegraphics[width=0.95\linewidth,height=0.8\linewidth,keepaspectratio]{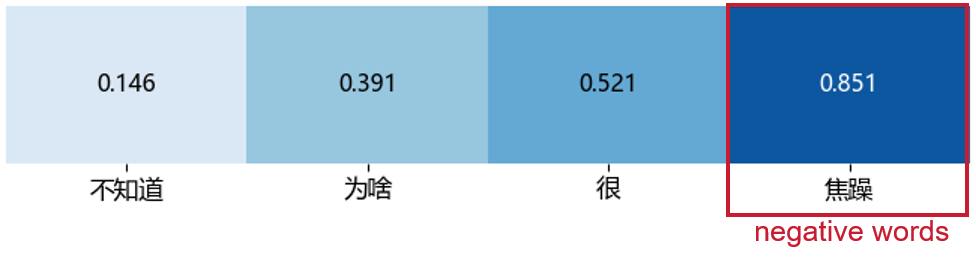}

    \vspace{-4mm}
    \caption{Visualization of gating weights in the ACMG module. The top curves and bottom heatmap illustrate the adaptive feature selection across acoustic and textual modalities, respectively. The blue shaded region denotes the low-energy region in the acoustic signal, while the red boxes highlight negative words in the transcript.}
    \label{fig:placeholder}
\end{figure}

As illustrated in Figure~\ref{fig:placeholder}, gating weights rise in low-energy regions (blue shaded areas) and are suppressed in high-energy segments, demonstrating an inverse relationship. These findings provide both quantitative and visual evidence that ACMG effectively emphasizes clinically relevant low-energy speech characteristics rather than operating in a random manner.

\subsubsection{Textual Gating Analysis} 
We further analyze the gating behavior on the textual modality. As illustrated in Figure \ref{fig:placeholder}, ACMG does not distribute weights uniformly across tokens; instead, it selectively amplifies linguistically salient cues. In particular, tokens expressing negative sentiments (highlighted by the red boxes in the heatmap) receive consistently higher gating scores. In contrast, semantically neutral or less informative tokens are relatively suppressed.This selective weighting pattern aligns with established psychiatric findings, which identify negative affect expressions and self-referential language as prominent linguistic markers of depression. The visualization thus provides qualitative evidence that the gating mechanism captures psychologically meaningful textual cues rather than operating in a content-agnostic manner.






\section{Conclusion}

In this work,we propose a multimodal depression detection network with Adaptive Cross-Modal Gating (ACMG) mechanism, which selectively highlights depression-related acoustic and textual segments. Experiments on PDCD2025 and DAIC-WOZ show that ACMG outperforms baseline models in both accuracy and F1. Visualization and ablation analyses confirm that it captures clinically meaningful patterns, including low-energy speech and negative sentiment text. Future work will explore alternative strategies for computing gating weights.





{
\section{Generative AI Use Disclosure}
Generative AI tools were used solely for manuscript language polishing. They were not used to create any core research content, results, or arguments. All authors are fully responsible for the work and consent to its submission. No generative AI tool is a co-author.}

\bibliographystyle{IEEEtran}
\bibliography{mybib}

\end{document}